\documentclass{iau}
\usepackage{graphicx}
\usepackage{amsmath}

\def\be{\begin{equation}}
\def\ee{\end{equation}}
\def\mr{\mathrm}
\def\dd{\mr{d}}
\def\hn{\hat{n}}
\def\xx{\mathbf{x}}
\def\ba#1\ea{\begin{align}#1\end{align}}
\def\lbra{\left\langle}
\def\rbra{\right\rangle}

\title[Combining cluster counts and galaxy clustering] 
{Combining cosmological constraints from cluster counts and galaxy clustering}

\author[Fabien Lacasa]   
{F. Lacasa $^1$}

\affiliation{$^1$ ICTP South American Institute for Research \& Instituto de F\'{i}sica Te\'{o}rica - UNESP \\  Rua Dr. Bento Teobaldo Ferraz 271, Bloco 2 - Barra Funda\\ 01140-070 S$\tilde{\mathrm{a}}$o Paulo, SP, Brazil \\ email: {\tt fabien@ift.unesp.br}}

\pubyear{2014}
\volume{xxx}  
\pagerange{}
\jname{Title of your IAU Symposium}
\editors{A.C. Editor, B.D. Editor \& C.E. Editor, eds.}

\begin{document}

\maketitle


\begin{abstract}
Present and future large scale surveys offer promising probes of cosmology. For example the Dark Energy Survey (DES) is forecast to detect $\sim$300 millions galaxies and thousands clusters up to redshift $\sim$1.3. I here show ongoing work to combine two probes of large scale structure : cluster number counts and galaxy 2-point function (in real or harmonic space). The halo model (coupled to a Halo Occupation Distribution) can be used to model the cross-covariance between these probes, and I introduce a diagrammatic method to compute easily the different terms involved. Furthermore, I compute the joint non-Gaussian likelihood, using the Gram-Charlier series. Then I show how to extend the methods of Bayesian hyperparameters to Poissonian distributions, in a first step to include them in this joint likelihood.
\end{abstract}


\firstsection 

\section{Cross-covariance between cluster counts and galaxy clustering}
\subsection{Cluster counts and galaxy angular 2-point function}
The number counts in a bin of mass $i_M$ and redshift $i_z$, can be considered as a monopole of the halo density field :
\be
\hat{N}_\mr{cl}(i_M,i_z) = \overline{N}_\mr{cl}(i_M,i_z) + \frac{1}{\Omega_S} \int \dd^2 \hn \, \dd M \, \dd z \; r^2 \frac{\dd r}{\dd z} \, \frac{\dd^2 n_h}{\dd M \dd V} \; \delta_\mr{cl}(\xx=r\hn | M,z)
\ee
Cluster counts have been shown as a powerful probe of cosmology, e.g. \cite{Planck2013-SZ} has produced constraint on $\sigma_8$ and $\Omega_m$ with SZ detected clusters.

 The study of the clustering of galaxies may be done with the angular correlation function 
$w(\theta)$ or its harmonic transform $C_\ell$, in tomographic redshift bins~:
\be
C_\ell^\mr{gal}(i_z,j_z) = \frac{2}{\pi} \int k^2 \dd k \, \frac{\overline{n}_\mr{gal}(z_1) \, \overline{n}_\mr{gal}(z_2) \, \dd V_1 \, \dd V_2}{\Delta N_\mr{gal}(i_z) \Delta N_\mr{gal}(j_z)} \, j_\ell(k r_1) \, j_\ell(k r_2) \; P_\mr{gal}(k | z_1,z_2)
\ee
In the following I use $C_\ell$ instead of $w(\theta)$ for simpler equations, although they can be related by a simple linear transformation.

\subsection{Cross-covariance derivation with the halo model}
\begin{figure}[htbp]
\begin{center}
\includegraphics[width=.7\linewidth]{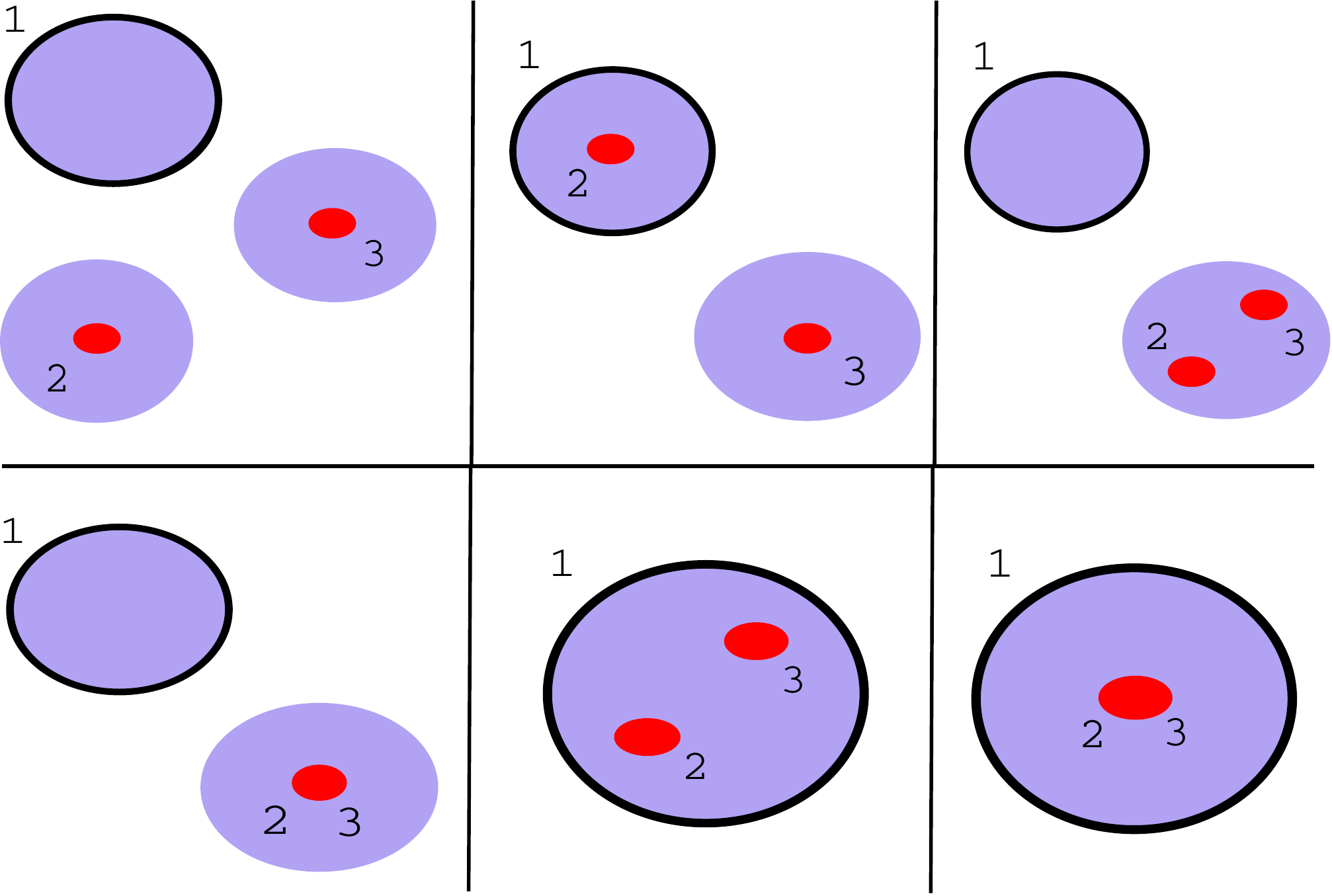}
\caption{Diagrams for the hgg bispectrum : 3h, 2h\_2h, 2h\_1h2g, 2h\_1h1g, 1h2g and 1h1g.
The 3h diagram has two contributions : non-linear evolution of dark matter (2PT), and second-order halo bias ($b_2$).}
\label{Fig:diagrams}
\end{center}
\end{figure}

The cross covariance between these two probes involves the halo-galaxy-galaxy angular bispectrum in the squeezed limit \cite{my-paper}~:
\be
\mr{Cov}\left(\hat{N}_\mr{cl}(i_M,i_z) , C_\ell^\mr{gal}(j_z, k_z)\right) = \int \frac{\dd M_1 \, \dd z_{123}}{4\pi} \, \frac{\dd V}{\dd z_1} \, \left.\frac{\dd^2 n_h}{\dd M \, \dd V}\right|_{M_1,z_1} b_{0\ell\ell}^{hgg}(M_1,z_{123})
\ee
$b_{0\ell\ell}$ is a projection of the 3D bispectrum, for which we need a non-linear model. In the framework of the halo model + HOD, I have shown that the bispectrum (or higher orders) can be computed with a diagrammatic formalism \cite{Lacasa2014}. See the diagrams for this hgg bispectrum in Fig. \ref{Fig:diagrams}.

\subsection{Current results}
I have shown that the equations for the covariance can be rewritten in terms of effective quantities, e.g. \cite{my-paper}~:
\ba
\nonumber \mr{Cov}_\mr{2PT}\left(\hat{N}_\mr{cl}(i_M,i_z) , C_\ell^\mr{gal}(j_z, k_z)\right) &= \frac{\delta_{j_z,k_z}}{4\pi} \int \frac{\overline{n}_\mr{gal}(z_2)^2 \, \dd V_1 \, \dd V_2}{\Delta N_\mr{gal}(j_z)^2} \, 4 F_\mr{sqz} \; b_1^\mr{gal,eff}(k_\ell,z_2)^2 \\
& \qquad \rho b_1^\mr{halo,eff}(i_M,z_1) \, P_\mr{DM}(k_\ell,z_2) \, \Delta_{0,P}(z_1,z_2)
\ea
These intermediate quantities are integrated over the halo mass and contain the HOD and mass function dependency. They can be compared to measurement on data or to other modeling, providing the possibility for some model independence.

\begin{figure}[htbp]
\begin{center}
\includegraphics[width=.7\linewidth]{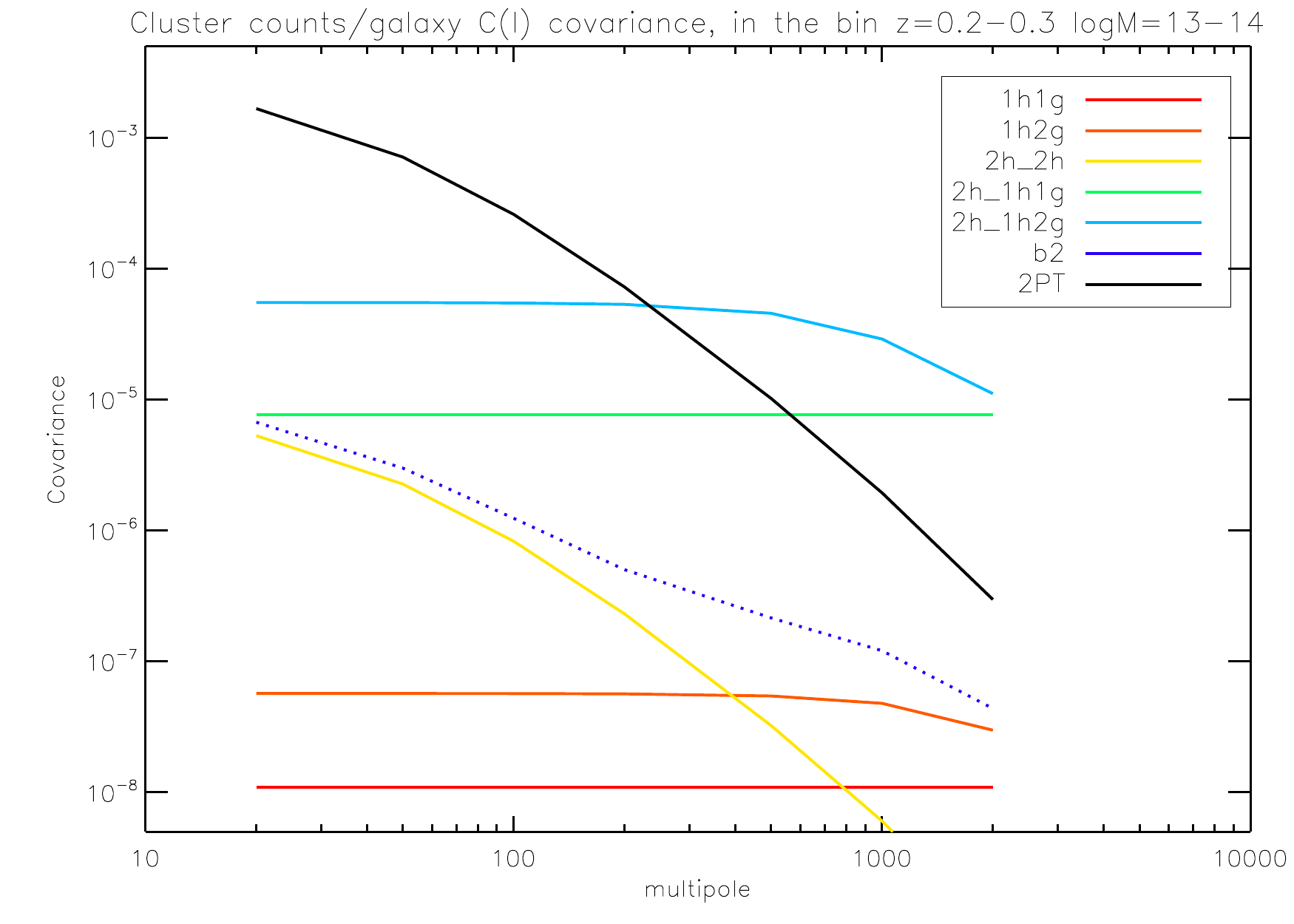}
\includegraphics[width=.7\linewidth]{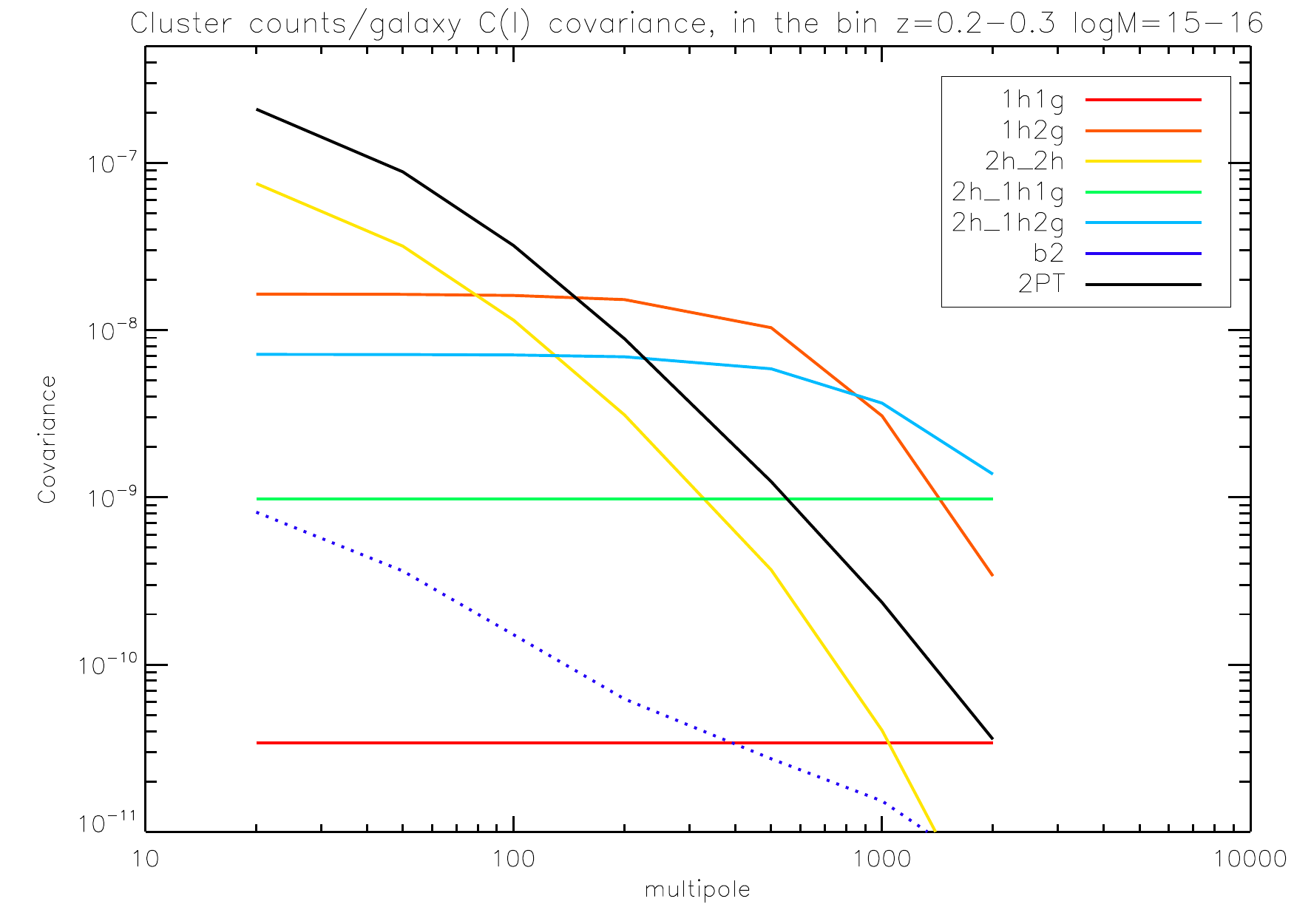}
\includegraphics[width=.7\linewidth]{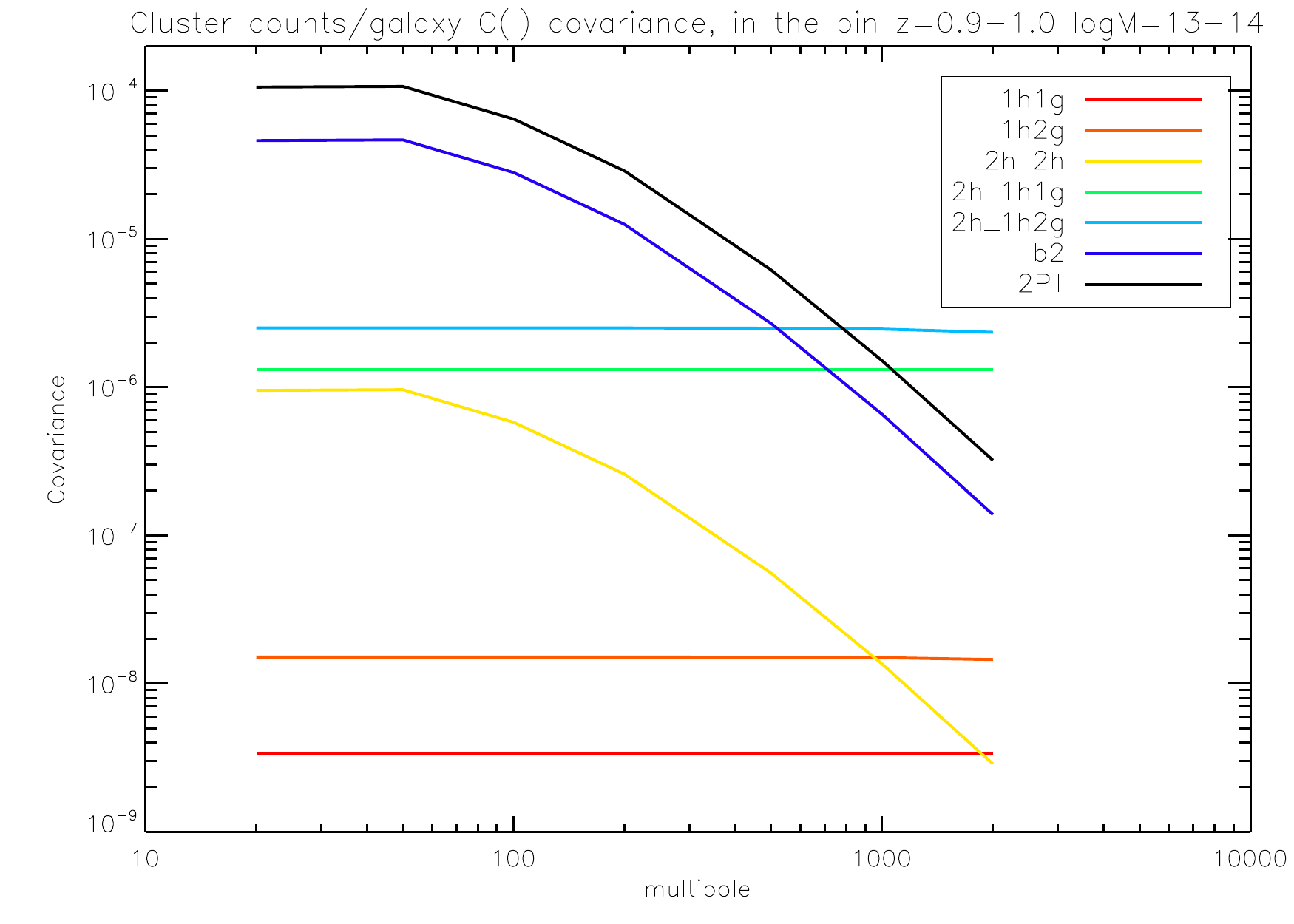}
\caption{Terms of the covariance for some bins of mass and redshift. {\bf From left to right~:} log$\,$M=13-14 \& z=0.2-0.3 ; log$\,$M=15-16 \& z=0.2-0.3 ; log$\,$M=13-14 \& z=0.9-1.
The $b_2$ term can be either negative (dotted line) at low $z$ when galaxies are antibiased,  or positive (solid line) at high $z$ when galaxies are biased.}
\label{Fig:Cov}
\end{center}
\end{figure}

I have built a fast and efficient code to compute the different terms of the covariance ; the plots in Fig. \ref{Fig:Cov} illustrate the numerical results. We see that different terms can become important depending on mass and redshift. The code runs in $\sim\!1$ CPU-second, and is thus adequate to be integrated into an MCMC pipeline.


\section{Likelihood}
\subsection{Joint non-Gaussian likelihood}\label{Sect:JointLikely}
Cluster counts follow a Poissonian distribution (up to sample variance), thus one cannot assume that the joint likelihood of $X=(\mr{counts},w_\mr{gal}(\theta))$ is Gaussian.\\
To tackle this, I expanded the joint likelihood with the Gram-Charlier series, around a fiducial independent case. I am then able to resum the expansion into \cite{my-paper}~:
\be
\mathcal{L}(X) = \exp\Big[-\sum_{i,j} \lbra c_{i} \, w_{j}\rbra_c \left(\log\lambda_{i} - \Psi(c_{i}+1)\right) (^T w C^{-1} e_{j})\Big] \ \mathcal{L}(\mr{counts}) \; \mathcal{L}(w)
\ee
This analytic form is well-behaved (positive), can be extended straightforwardly to include sample variance, and has correct asymptotic behaviour at large $N_\mr{cl}$ (Gaussian with the correct covariance matrix).

\subsection{Bayesian hyperparameters}
\begin{figure}[htbp]
\begin{center}
\includegraphics[width=.7\linewidth]{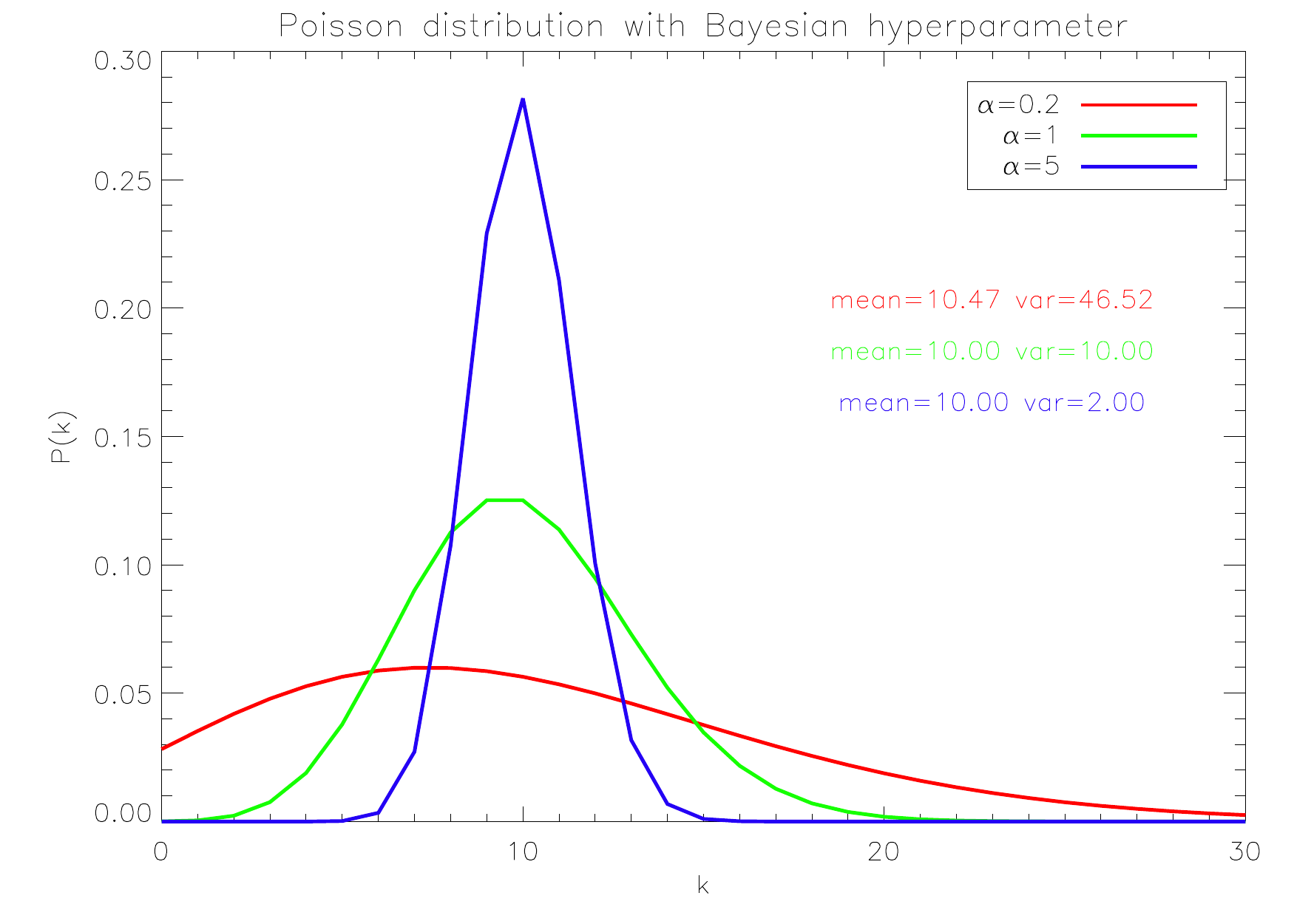}
\caption{Diagrams for the hgg bispectrum : 3h, 2h\_2h, 2h\_1h2g, 2h\_1h1g, 1h2g and 1h1g.
The 3h diagram has two contributions : non-linear evolution of dark matter (2PT), and second-order halo bias ($b_2$).}
\label{Fig:PoissonHP}
\end{center}
\end{figure}

Hyperparameters (HPs) allow to detect over/underestimation of error bars, or inconsistencies between data sets \cite{Hobson2002}.
The method is at the moment only adapted to Gaussian distributions, thus not for Poissonian cluster counts.
It is however mathematically impossible to keep the Gaussian properties of HPs in the Poissonian case (that is, rescaling the variance while keeping the mean). However I found a prescription which approximately respect them. On Fig. \ref{Fig:PoissonHP} are shown three pdfs, corresponding to three different values of the Bayesian HP $\alpha$.\\
This will allow the use of HPs for the cluster counts - galaxy 2-pt combination, after further extension (sample variance, correlation with $w_\mr{gal}$ as treated in Sect. \ref{Sect:JointLikely}).


\section{Conclusion and perspectives}
I sketched how to combine cluster counts and galaxy 2-pt measurements for increased cosmological constraints, from the physical modeling to the likelihood. In the context of the halo model, I introduced a diagrammatic method allowing elegant computation of the equations involved. I derived a non-Gaussian joint likelihood using Gram-Charlier series, and showed how to introduce Bayesian HPs to a Poissonian distribution.

Further work will be necessary to include experimental effects in the covariance and the likelihood : photo-z errors, purity... Next order derivation of the joint likelihood may also be necessary to solve  a small bias issue at low $N_\mr{cl}$, and the bayesian hyperparameters method need to be extended to cluster sample variance and correlation with galaxies. In the medium term, I aim to build a full MCMC pipeline of cluster-galaxy combination, for realistic forecasts and application to DES data.

Further details on the model, method, and forecasts will be available in Lacasa \& Rosenfeld (in prep.)


\section{Figures}

\end{document}